\documentclass[
reprint,
 amsmath,amssymb,
 aps,
pre,
showpacs
]{revtex4-1}

\usepackage{graphicx}
\usepackage{dcolumn}
\usepackage{bm}

\usepackage{subfigure}
\usepackage[bookmarks=false,pdfstartview=FitH,hyperindex=true, colorlinks, linkcolor=blue, citecolor=green,anchorcolor=red]{hyperref}
\usepackage{comment}
\usepackage{listings}
\usepackage{algpseudocode}
\usepackage{algorithm}
\usepackage{epstopdf}
\usepackage{hyperref}
\usepackage{float}
\usepackage{pgf}
\usepackage{tikz}

\newcommand{\overbar}[1]{\mkern 1.5mu\overline{\mkern-1.5mu#1\mkern-1.5mu}\mkern 1.5mu}

\begin{document}

\title{Lifted Worm Algorithm for the Ising Model}

\author{Eren Metin El\c{c}i$^1$}
\author{Jens Grimm$^2$}
\author{Lijie Ding$^3$}
\author{Abrahim Nasrawi$^2$}
\author{Timothy M. Garoni$^2$}
\author{Youjin Deng$^3$}
\affiliation{$^1$School of Mathematical Sciences, Monash University, Clayton, Victoria 3800, Australia}
\affiliation{$^2$ARC Centre of Excellence for Mathematical and Statistical Frontiers
(ACEMS), School of Mathematical Sciences, Monash University, Clayton,
VIC 3800, Australia}
\affiliation{$^3$Department of Physics, University of Science and Technology of China, Hefei, Anhui 230026, China}
\affiliation{$^4$National Laboratory for Physical Sciences at Microscale and Department of Modern Physics, University of Science and Technology of China, Hefei, Anhui 230026, China}


\begin{abstract}
We design an irreversible worm algorithm for the zero-field ferromagnetic Ising model by using the lifting technique. We study the dynamic critical behavior of an energy estimator on both the complete graph and toroidal grids, and compare our findings with reversible algorithms such as the Prokof'ev-Svistunov worm algorithm. Our results show that the lifted worm algorithm improves the dynamic exponent of the energy estimator on the complete graph, and leads to a significant constant improvement on toroidal grids.
\end{abstract}

\maketitle


\section{Introduction} 
\label{sec:introduction}
Markov-chain Monte Carlo (MCMC) algorithms are a powerful and widely-used tool in various areas of physics and other disciplines, such as in machine learning~\cite{2016_goodfellow} and statistics~\cite{2016_efron}. In many practical applications MCMC algorithms are constructed via the Metropolis~\cite{1953_metropolis} or heat bath update scheme ~\cite{2008_janke}. Such algorithms are necessarily \textit{reversible}.\par
One important example of a Metropolis algorithm is the Prokof'ev-Svistunov Worm Algorithm (P-S worm algorithm) which has widespread application for both classical and quantum systems~\cite{2001_prokofev,2006_prokofev}. As opposed to cluster algorithms like the Wolff~\cite{1989_wolff} or Swendsen-Wang algorithm~\cite{1987_swendsen}, the updates of the worm algorithm are purely \textit{local}. On the simple-cubic lattice with periodic boundaries, it was numerically observed that the P-S worm algorithm for the zero-field ferromagnetic Ising model outperforms the Swendsen-Wang algorithm for simulating both the magnetic susceptibility and the second-moment correlation length~\cite{2007_deng}. Another numerical work suggested that the spin-spin correlation function can also be simulated efficiently~\cite{2009_wolff}. Recently, it was rigorously established~\cite{2016_colle} that the P-S worm algorithm for the Ising model is rapidly mixing on any finite graph for the whole temperature range.\par
In recent years, various \textit{irreversible} MCMC algorithms have also been studied~\cite{2010_hayes,2011_turitsyn,2010_suwa,2011_fernandes,2009_bernard, 2013_sakai, 2013_engel, 2015_kapfer, 2015_michel, 2015_nishikawa, 2017_hu}. Many of these algorithms are based on the lifting technique introduced in~\cite{2001_diaconis}. The general idea of lifting is to enlarge the original state space and define transition probabilities such that the lifted chain projects down to the original one. The intuition underlying a potential efficiency improvement is the reduction of diffusive behavior, compared with the original Markov-chain. Rather than exploring states via random walk in the reversible chain one introduces directed flows in the lifted chain to move between relevant states significantly faster.\par 
Even though lifting is considered as a promising method to speed up MCMC algorithms, it is an open question how it affects efficiency in specific examples~\cite{2013_diaconis}. For the Ising model on the complete graph, it was numerically observed that the lifted single-spin flip Metropolis algorithm improves the scaling (with volume) of the rate of decay of the autocorrelation function of the magnetization~\cite{2011_turitsyn}. 
Another study~\cite{2010_hayes} proved that a lifted MCMC algorithm for uniformly sampling leaves from a given tree reduces the mixing time. In other examples~\cite{2015_michel,2011_fernandes,2017_hu} it was numerically observed that lifting speeds up reversible MCMC algorithms by a possibly large constant factor but does not asymptotically affect the scaling with the system size.\par
In this work we investigate how lifting affects worm algorithms. More precisely, we design a lifted worm algorithm for the zero-field ferromagnetic Ising model, and numerically study the dynamic critical behavior of an estimator of the energy. Our simulations were performed on both the complete graph and toroidal grids in dimensions $2\leq d \leq 5$ at the (estimated, when $d \ge 3$) infinite-volume critical point. \par
On the complete graph we find that the lifted worm algorithm significantly improves the dynamic critical exponent $z_{\rm int}$ of the energy estimator. In particular, we show that the energy estimator exhibits critical speeding-up~\cite{2007_deng_2,2011_pomeau} in the lifted process ($z_{\rm int} \approx -0.5$), while we observe $z_{\rm int} \approx 0$ for the corresponding reversible counterpart. On toroidal grids we find that the lifted worm algorithm does not affect the scaling with the system size. We emphasize, however, that the lifted process still reduces the variance of the energy estimator by a significant constant. This constant improvement becomes more pronounced for larger dimensions with up to a factor of approximately $141$ for $d=5$. We, finally, note that the improvements in high dimensions can be of practical relevance in the current debate about the correct finite-size scaling behavior above the upper critical dimension, see e.g.~\cite{2014_lundow,2014_wittmann,2016_flora,2017_grimm}.\par
This paper is organized as follows. Section~\ref{sec:reversible_algorithm} recalls the basic ideas of the P-S worm algorithm. In Sec.~\ref{sec:irreversible_algorithm} we construct an irreversible worm algorithm via the lifting technique. We present the details of our numerical setup in Sec.~\ref{sec:numerical_setup}. In Sec.~\ref{sec:results} we state our results for the dynamic properties of the lifted worm algorithm, and compare our findings with reversible worm algorithms. Finally, in Sec.~\ref{sec:discussion} we summarize and discuss our findings.

\section{P-S Worm Algorithm} 
\label{sec:reversible_algorithm}
As is well known~\cite{1979_thompson}, the zero-field ferromagnetic Ising model can be mapped to an ensemble of high-temperature graphs. Let $G=(V,E)$ be a finite graph with vertex set $V$ and edge set $E$. Define the closed loop space $\mathcal{C}_0$ as the set of all configurations $\omega \subseteq E$ such that every vertex has even degree, and $\mathcal{C}_2$ as the set of all $\omega \subseteq E$ where exactly two vertices have odd degree. We call the subgraph $(V, \omega)$ Eulerian whenever $\omega \in \mathcal{C}_0$. In the high-temperature expansion the partition function of the Ising model can be written as the sum over all Eulerian subgraphs \cite{1979_thompson}, ~i.e.
    \begin{equation}
    \label{eq:worm_partition_function}
    Z = 2^{\left|V\right|}\cosh^{\left|E\right|}(\beta) \sum_{\omega\in \mathcal{C}_0} \tanh^{\left|\omega\right|}(\beta),
    \end{equation}
where $\beta$ denotes the inverse Ising temperature, and $\left | \cdot \right|$ denotes the cardinality of the corresponding set. \par 
The P-S worm algorithm samples these high-temperature graphs via elementary local moves. The main idea is to enlarge the state space $\mathcal{C}_0$ to $\mathcal{W}: = \mathcal{C}_0 \cup \mathcal{C}_2$ by introducing two vertices with odd degree (defects). These defects are moved through $\mathcal{W}$ via random walk. Whenever the two defects meet, the subgraph becomes Eulerian and one reaches a state of the original configuration space $\mathcal{C}_0$.\par
In the original algorithm \cite{2001_prokofev} only one of the defects is mobile and can be moved through $\mathcal{W}$. In this work, we use a slightly different worm version where we flip a fair coin to select the mobile defect in $\mathcal{C}_2$. The \textit{transition probabilities}
\begin{equation}
q(\omega,\omega^\prime)=p(\omega,\omega^\prime)a(\omega,\omega^\prime)
\label{eq:transition_matrix}
\end{equation}
from an initial state $\omega$ to a target state $\omega^\prime$ can be calculated by metropolizing~\cite{1953_metropolis} the proposals $p(\omega,\omega^\prime)$ with respect to the stationary measure $\pi(\omega)= \frac{1}{\mathcal{Z}} \tanh^{\left|\omega\right|}(\beta) \Psi(\omega)$. Here, $\mathcal{Z}$ is an appropriate normalization on $\mathcal{W}$, $\Psi(\omega):=\left|V\right|$ if $\omega \in \mathcal{C}_0$, and $\Psi(\omega):=2$ if $\omega \in \mathcal{C}_2$, respectively. Furthermore, $p(\omega,\omega^\prime)$ denotes the \textit{proposal probability} and $a(\omega,\omega^\prime)$ the \textit{acceptance probability}. See~\cite{2016_colle} for explicit expressions for the transition matrix~\eqref{eq:transition_matrix}.\par
The algorithm is presented in Alg.~\ref{alg:rev_worm} with $\omega\Delta xx^\prime$ denoting the symmetric difference of $\omega$ and the edge $xx^\prime$; i.e.~ $xx^\prime \in \omega \Delta xx^\prime$ if and only if $xx^\prime \notin \omega$.
\begin{algorithm}[H]
\caption{P-S Worm Algorithm}
\label{alg:rev_worm}
\begin{algorithmic}

    \If {$\omega \in \mathcal{C}_0$}
    \State 
		Choose a uniformly random vertex $x$
    \Else \State
        Choose a uniformly random odd vertex $x$
    \EndIf
    \State
    Choose a uniformly random edge $xx^\prime$ among the set of edges incident to $x$. With probability $a_{\text{P-S}}(\omega, \omega \Delta xx^\prime)$, let $\omega \to \omega \Delta xx^\prime$. Otherwise $\omega \to \omega$ 
    \end{algorithmic}
    \end{algorithm}
\medskip

\section{Irreversible Worm Algorithm} 
\label{sec:irreversible_algorithm}

We construct the irreversible worm algorithm in two steps. In Sec.~\ref{sec:bs_worm} we first define an alternative \textit{reversible} worm algorithm. This worm algorithm will be an appropriate starting point to apply lifting. In Sec.~\ref{sec:irre_worm} we use lifting to construct the irreversible counterpart.

\subsection{B-S type Worm Algorithm}
\label{sec:bs_worm}
Since it is not obvious how to apply the lifting technique to the P-S process in a natural way, we first construct an alternative reversible worm algorithm with slightly different proposals. This algorithm can be seen as the Ising analogue of the Berretti-Sokal algorithm \cite{1985_berretti} for simulating self-avoiding walks in the grand canonical ensemble. We thus call it the \textit{B-S type worm algorithm}. \par
The proposals are as follows: We first decide to either increase $(+)$ or decrease $(-)$ the number of occupied edges by flipping a fair coin. Then, if the current state belongs to $\mathcal{C}_2$, we flip a fair coin to select one of the two defects as the mobile vertex. Otherwise, if the current state is an element of $\mathcal{C}_0$, we choose a uniformly random vertex as the mobile vertex. If we decide to add (delete) an edge, we  select the next position of the mobile vertex uniformly at random among the set of vacant (occupied) edges incident to the current mobile vertex. We now construct the transition probabilities by metropolizing the proposals with respect to the same measure as in Sec.~\ref{sec:reversible_algorithm}.\par
For $\omega\in\mathcal{W}$, $v\in V$ and $\lambda\in\{-1,1\}$, define
\begin{align}
N_{\omega}(x,\lambda)=\left\{\begin{array}{ll}
\{uv\not\in\omega\;:\;u=x\text{ or }v=x\}, &\text{if } \lambda=+1\\
\{uv\in\omega\;:\;u=x\text{ or }v=x\}, &\text{if } \lambda=-1
\end{array}\right.
\end{align}
Note that, for any $\omega\in\mathcal{W}$, $N_{\omega}(x,+1)+N_{\omega}(x,-1)$ equals the degree of $x$.\par
Fix $z:=\tanh(\beta)$, and let $\omega,\omega\Delta xx'\in\mathcal{W}$. The proposal and acceptance probabilities for the transition $\omega\to\omega\Delta xx'$ are as follows \medskip
\begin{widetext}
(i) If $\omega \in \mathcal{C}_0$:
\begin{equation}
p_{\text{B-S}}(\omega,\omega \Delta xx^\prime)= \frac{1}{2} \frac{1}{\left|V\right|} \Bigg [\frac{1}{|N_{\omega}(x,|\omega\Delta xx'|-|\omega|)|}+\frac{1}{|N_{\omega}(x',|\omega\Delta xx'|-|\omega|)|}\Bigg ]
\end{equation}
\begin{equation}
a_{\text{B-S}}(\omega,\omega \Delta xx^\prime)= \min \Bigg [ 1,\frac{z^{|\omega\Delta xx'|}}{z^{|\omega|}}\frac{|N_{\omega\Delta xx'}(x,|\omega|-|\omega\Delta xx'|)|^{-1}+|N_{\omega\Delta xx'}(x',|\omega|-|\omega\Delta xx'|)|^{-1}}{|N_{\omega}(x,|\omega\Delta xx'|-|\omega|)|^{-1}+|N_{\omega}(x^\prime,|\omega\Delta xx'|-|\omega|)|^{-1}} \Bigg ]
\end{equation}

\medskip

(ii) If $\omega \in \mathcal{C}_2$, $\omega \Delta xx^\prime \in \mathcal{C}_2$ and $x$ is a defect in $\omega$:
\begin{equation}
p_{\text{B-S}}(\omega,\omega \Delta xx^\prime) = \frac{1}{2}\frac{1}{2}\frac{1}{|N_{\omega}(x,|\omega\Delta xx'|-|\omega|)|}
\label{eq:proposal_figure}
\end{equation}
\begin{equation}
a_{\text{B-S}}(\omega,\omega \Delta xx^\prime)= \min \Bigg [ 1,\frac{z^{|\omega\Delta xx'|}}{z^{|\omega|}}\frac{|N_{\omega}(x,|\omega\Delta xx'|-|\omega|)|}{|N_{\omega\Delta xx'}(x,|\omega|-|\omega\Delta xx'|)|}\Bigg ]
\label{eq:acceptance_figure}
\end{equation}

(iii) If $\omega \in \mathcal{C}_2$, $\omega \Delta xx^\prime \in \mathcal{C}_0$:
\begin{equation}
p_{\text{B-S}}(\omega,\omega \Delta xx^\prime)= \frac{1}{2}\frac{1}{2}\Bigg [\frac{1}{|N_{\omega}(x,|\omega\Delta xx'|-|\omega|)|}+\frac{1}{|N_{\omega}(x',|\omega\Delta xx'|-|\omega|)|}\Bigg ]
\end{equation}
\begin{equation}
a_{\text{B-S}}(\omega,\omega \Delta xx^\prime)=\min \Bigg [ 1,\frac{z^{|\omega\Delta xx'|}}{z^{|\omega|}}\frac{|N_{\omega\Delta xx'}(x,|\omega|-|\omega\Delta xx'|)|^{-1}+|N_{\omega\Delta xx'}(x',|\omega|-|\omega\Delta xx'|)|^{-1}}{|N_{\omega}(x,|\omega\Delta xx'|-|\omega|)|^{-1}+|N_{\omega}(x^\prime,|\omega\Delta xx'|-|\omega|)|^{-1}} \Bigg ]
\end{equation}\medskip
\end{widetext}

\begin{figure}
\includegraphics[scale=0.5]{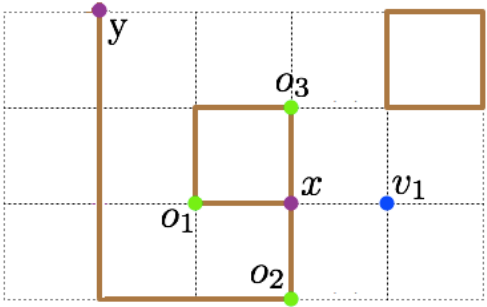}
\caption{Worm configuration $\omega \in \mathcal{C}_2$ where $x,y$ are the vertices with odd degree, $N_{\omega}(x,-)=\{xo_1,xo_2,xo_3\}$, and $N_{\omega}(x,+)=\{xv_1\}$. If $x$ is selected as the mobile vertex and one proposes to increase the number of edges in the \textit{B-S type worm algorithm}, the only possible transition is $\omega \to \omega \Delta xv_1$ where $\omega \Delta xv_1 \in \mathcal{C}_2$. The corresponding proposal and acceptance probabilities are stated in Eq.~\eqref{eq:proposal_figure} and~\eqref{eq:acceptance_figure}.
} 
\label{fig:worm_config}
\end{figure}

We give an example for case (ii) in Fig.~\ref{fig:worm_config}. All other off-diagonal transition probabilities are zero. The full algorithmic description is presented in Alg.~\ref{alg:bs_worm}. We remark that the choice to allow both defects to move in Alg.~\ref{alg:bs_worm} is not actually necessary, and one can construct a modification of Alg.~\ref{alg:bs_worm} in which only one defect is mobile.

\begin{algorithm}[H]
    \caption{B-S type Worm Algorithm} \label{alg:bs_worm}
    \begin{algorithmic}
    	\State
     		Choose $\lambda=\{+,-\}$ uniformly at random
    \If {$\omega \in \mathcal{C}_0$}
    \State 
		Choose a uniformly random vertex $x$
    \Else \State
        Choose a uniformly random odd vertex $x$
    \EndIf
    \State
    \If {$N_{\omega}(x,\lambda)=\emptyset$}
    \State 
    	Set $\omega\to\omega$ and skip all following steps
    	\Else \State
    	Choose a uniformly random edge $xx'\in N_{\omega}(x,\lambda)$. With probability $a_{\text{B-S}}(\omega, \omega \Delta xx^\prime)$, let $\omega \to \omega \Delta xx^\prime$.  Otherwise $\omega \to \omega$.
	\EndIf
    \end{algorithmic}
    \end{algorithm}

\subsection{Irreversible Worm Algorithm}
\label{sec:irre_worm}
In the following, we construct the irreversible counterpart of the B-S type worm algorithm. Consider the enlarged state space $\mathcal{W}^\prime:= \mathcal{W} \times \{-,+\}$ where $\{-,+\}$ is a set to indicate to either choose to increase $(+)$ or decrease $(-)$ the number of edges. Our aim is to define a Markov-chain on $\mathcal{W}^\prime$ such that we never propose to delete an edge if a state belongs to $\mathcal{W}\times \{+\}$, while we never propose to add edges as long as the chain belongs to $\mathcal{W} \times \{-\}$. If a move $(\omega,\lambda)\to(\omega\Delta xx',\lambda)$ is rejected, we make the transition $(\omega,\lambda)\to(\omega,-\lambda)$. Note that this process does not allow diagonal transitions.\par
For $(\omega,\lambda)\in\mathcal{W'}$, let $\tilde{\pi}(\omega,\lambda)=\frac{1}{2}\pi(\omega)$. For $xx'\in E$, let
\begin{align}
\tilde{q}((\omega,+),(\omega\cup xx',+))&=q_{\text{B-S}}(\omega,\omega\cup xx')\text{  if  }xx'\notin\omega\\
\tilde{q}((\omega,-),(\omega\setminus xx',-))&=q_{\text{B-S}}(\omega,\omega\setminus xx')\text{  if  }xx'\in\omega
\end{align}
All other entries in row $\tilde{q}((\omega,\lambda),(\cdot,\cdot))$ are zero except $\tilde{q}((\omega,\lambda),(\omega,-\lambda))$ which is fixed by stochasticity. Observe that skew-detailed balance~\cite{2011_turitsyn,2016_vucelja} between $\tilde{q}$ and $\tilde{\pi}$ follows immediately from detailed balance between $q$ and $\pi$, and so $\tilde{q}$ has stationary distribution $\tilde{\pi}$.\par
The full algorithmic description of the lifted worm algorithm is given in Alg.~\ref{alg:irre_worm}. From a practical perspective, we emphasize that only minor code changes to the B-S type worm algorithm are needed to construct the irreversible counterpart.

    \begin{algorithm}[H]
    \caption{Irreversible Worm Algorithm}
    \label{alg:irre_worm}
    \begin{algorithmic}
    \If {$\tilde{\omega}=(\omega,\lambda)$ where $\omega \in \mathcal{C}_0$}
    \State 
		Choose a uniformly random vertex $x$
        \Else\State
        Choose a uniformly random odd vertex $x$
    \EndIf
    \State
    \If{$N_{\omega}(x,\lambda) = \emptyset$}
    \State Set $(\omega,\lambda) \to (\omega,-\lambda)$ and skip all following steps
    \Else\State
    Choose a uniform random edge $xx'\in N_{\omega}(x,\lambda)$. With probability $a_{\text{B-S}}\big(\omega, \omega \Delta xx^\prime\big)$, let $(\omega,\lambda) \to (\omega\Delta xx^\prime,\lambda)$.  Otherwise $(\omega,\lambda) \to (\omega,-\lambda)$
\EndIf
\State
    \end{algorithmic}
    \end{algorithm}


\section{Numerical setup}
\label{sec:numerical_setup}
Let $\omega \in \mathcal{W}$. We numerically study dynamic properties of the \textit{number of occupied edges} $\mathcal{N}:=\left|\omega\right |$ where $\langle \mathcal{N} \rangle$ is related to the energy when measurements are taken in the Eulerian subspace~\cite{2001_prokofev}. $\langle \mathcal{N} \rangle$ denotes the ensemble average estimated via the arithmetic mean $\overbar{\mathcal{N}}=\frac{1}{M}\sum_{i=1}^M \mathcal{N}_i$ where $M$ is total number of measurements, and $\mathcal{N}_i$ the value of the random variable at the $i$-th Monte Carlo step after a sufficiently long burn-in sequence has been discarded.\par
In Sec.~\ref{sec:results} we compare the variance of $\overbar{\mathcal{N}}$ (in the limit of $M\rightarrow \infty$) by using
\begin{equation}
\text{Var}(\overbar{\mathcal{N}}) \sim 2\tau_{\text{int}}^{(\mathcal{N})} \frac{\text{Var}(\mathcal{N}_0)}{M} \quad M\rightarrow \infty
\label{eq:def_var}
\end{equation}
among the P-S, B-S type and irreversible worm algorithms. Here, $\tau_{\text{int}}^{(\mathcal{N})}$ is the \textit{integrated autocorrelation time}
\begin{equation}
\tau_{\text{int}}^{(\mathcal{N})}:= \frac{1}{2}+\sum_{t=1}^{\infty} \rho^{(\mathcal{N})}(t).
\label{eq:def_tau_int}
\end{equation}
where $\rho^{(\mathcal{N})}(t)$ denotes the \textit{normalized autocorrelation function}\par
\begin{equation}
\rho^{(\mathcal{N})}(t): = \frac{\langle \mathcal{N}_0 \mathcal{N}_{t} \rangle - \langle \mathcal{N}_0 \rangle^2}{\text{Var}(\mathcal{N}_0)}.
\end{equation}
Our simulations for the Ising model were performed on the complete graph on $n$ vertices, and on toroidal grids for $2 \le d \le 5$. On the complete graph we simulated at the critical point $\beta_{\rm crit} = 1/n$. On the torus our simulations were performed at the exact critical point in two dimensions~\cite{1982_baxter}, and at the estimated critical points $\beta_{\text{crit},3d} = 0.22165455(3)$~\cite{2003_deng}, $\beta_{\text{crit},4d} = 0.1496947(5)$~\cite{2012_lundow}, and $\beta_{\text{crit},5d} = 0.1139150(4)$~\cite{2014_lundow} for $d\ge 3$.\par 
In each time series, we truncated the summation in Eq.~\eqref{eq:def_tau_int} self-consistently by using the windowing method~\cite{1996_sokal}. We emphasize that particular care has to be taken when choosing the windowing parameter $c$ for the irreversible worm algorithm, see Sec.~\ref{sec:results} and Appendix~\ref{ap:sokal}. For fitting and error estimations we follow standard procedures, see e.g.~\cite{1996_sokal, 2015_young}.\medskip
\section{Results} 
\label{sec:results}    
\subsection{Toroidal grids}
\label{subsec:toroidal_grids}
We will now study the dynamic properties of $\mathcal{N}$ on $d$ dimensional toroidal grids. Note that $\text{Var}(\mathcal{N}_0)$ in Eq.~\eqref{eq:def_var} coincides among all studied algorithms, since it is a property of the stationary measure and does not depend on the details of the underlying Markov-chain. Therefore, in the limit $M\rightarrow \infty$, we have $\frac{\text{Var}_{i}(\overbar{\mathcal{N}})}{\text{Var}_{j}(\overbar{\mathcal{N}})} = \frac{\tau_{\text{int},i}^{(\mathcal{N})}}{\tau_{\text{int},j}^{(\mathcal{N})}}$ where $i,j \in \{\text{P-S, B-S type, irre}\}$. \par
\begin{figure}[htb]
            \centering
\includegraphics[scale=0.45]{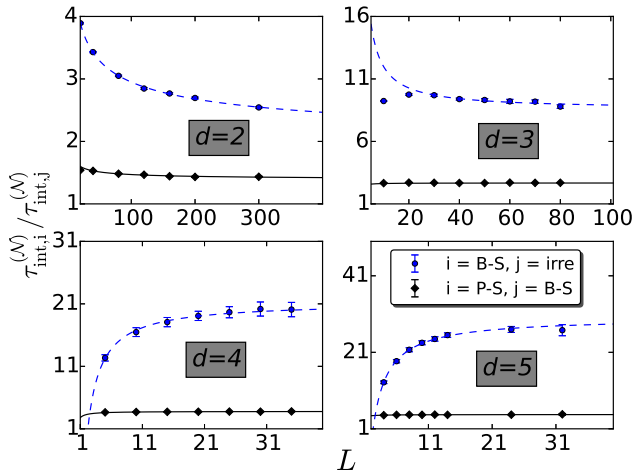}
\caption{Comparison of the integrated autocorrelation time among the P-S, B-S type and lifted worm algorithm. The black diamonds show the ratio of the integrated autocorrelation time of the P-S and B-S type algorithm, while the blue circles compare $\tau_{\rm int}^{(\mathcal{N})}$ among the lifted and the B-S type worm algorithm. The asymptotic improvement factors can be found in Table~\ref{table:torus_improvements}. The lines correspond to the fits in Sec.~\ref{subsec:toroidal_grids}.}
\label{fig:var_comparison_bond}
\end{figure}

In Fig.~\ref{fig:var_comparison_bond} we compare the integrated autocorrelation time among the B-S type and lifted worm algorithm (resp. B-S type and P-S worm algorithm). We perform least square fits of the form $A+BL^{-\Delta}$ where $A,B,\Delta$ are free parameters. Our conclusion will be that both ratios are approaching constants for $L \to \infty$, with larger improvements for higher dimensions, see Table~\ref{table:torus_improvements}. In two dimensions, our fits lead to the constant improvement $A_{\text{B-S}\to\text{irre}}=1.7(2)$, and $A_{\text{P-S}\to\text{B-S}}=1.4(1)$ by discarding $L<40$. For $d=3$, we find $A_{\text{B-S}\to\text{irre}}=8.2(4)$ by discarding $L<40$, and $A_{\text{P-S}\to\text{B-S}}=2.68(9)$ by discarding $L<20$. In four dimensions our fits lead to $A_{\text{B-S}\to\text{irre}}=24(1)$, and $A_{\text{P-S}\to\text{B-S}}=3.79(3)$ by discarding $L<10$. For $d=5$ we find $A_{\text{B-S}\to\text{irre}}=30(1)$, and $A_{\text{P-S}\to\text{B-S}}=4.7(1)$. In order to obtain stable fits we fixed $\Delta=1$ for fitting the ratios of the integrated autocorrelation time of the B-S type and lifted worm algorithm for $d>2$, and for fitting the ratios of the P-S and B-S type worm algorithm in four and five dimensions.\par
\begin{figure}[htb]
            \centering
\includegraphics[scale=0.35]{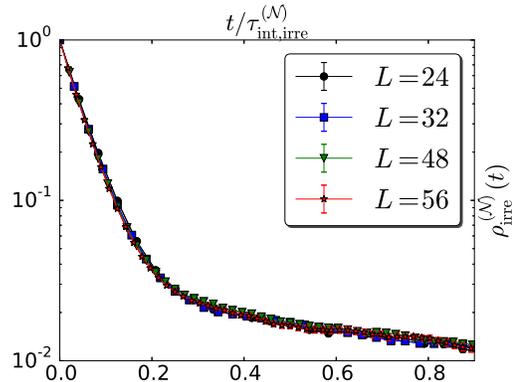}
\caption{Normalized autocorrelation function $\rho_{\rm irre}^{(\mathcal{N})}(t)$ ($t$ in MC hits) for the irreversible worm algorithm in five dimensions. As stated in Sec.~\ref{subsec:toroidal_grids}, $\rho_{\rm irre}^{(\mathcal{N})}(t)$ is well described by the ansatz in Eq.~\eqref{ansatz_auto_function_torus}.}
\label{fig:int_autocorrelation_time_comparison_bond}
\end{figure}

For estimating $\tau_{\text{int,irre}}^{(\mathcal{N})}$ in four and five dimensions we had to choose very large $c$ values outside the common range $c \in [6,10]$ in the windowing algorithm~\cite{1996_sokal} (see Appendix~\ref{ap:sokal} for details). In order to understand this, it proves useful to study the autocorrelation function $\rho_{\rm irre}^{(\mathcal{N})}(t)$ where $t$ is measured in MC hits. For clarity, we will only focus on the five dimensional case. Figure~\ref{fig:int_autocorrelation_time_comparison_bond} shows $\rho_{\rm irre}^{(\mathcal{N})}(t)$ in the lifted worm algorithm for $d=5$. $\rho_{\rm irre}^{(\mathcal{N})}(t)$ exhibits a two-time scaling: For small $t$, $\rho_{\rm irre}^{(\mathcal{N})}(t)$ shows a quick exponential decay to a small but bounded value while we observe a much slower decay with a different (larger) exponential scale for larger $t$. We note that this two-time scaling is absent for $d\le3$. Our data suggests that $\rho_{\rm irre}^{(\mathcal{N})}(t)$ can be described by the ansatz
\begin{equation}
\rho_{\rm irre}^{(\mathcal{N})}(t) = \alpha_1 \exp(-t/ \tau_{1})+ \alpha_2 \exp(-t/\tau_{2})
\label{ansatz_auto_function_torus}
\end{equation}
where $\alpha_1 + \alpha_2 = 1$, $\tau_{1}\geq\tau_{2}>0$.
In order to estimate the optimal least-squares parameters for this ansatz we used an improved procedure described in the Appendix~\ref{ap:appenfit}. We find that both the ratio $\alpha_1/\alpha_2$ and $\tau_2/\tau_1$ remain bounded as $L\rightarrow \infty$. We estimate the corresponding asymptotic constants using a least squares fitting procedure with a constant and obtain
$$
\frac{\alpha_2}{\alpha_1} \stackrel{L\rightarrow \infty}{\sim} 40.6(4) \quad \& \quad
\frac{\tau_1}{\tau_2} \stackrel{L\rightarrow \infty}{\sim} 32.4(6).
$$
Our numerical observation that $\alpha_1$ ($\tau_2$) is significantly smaller than $\alpha_2$ ($\tau_1$) impacts the required choice of parameters in windowing algorithm~\cite{1996_sokal}.


\begin{table}
\centering
\begin{tabular}{c|c|c|c} 
• & P-S $\to$ B-S & B-S $\to$ irre & P-S $\to$ irre \\ 
\hline
$d=2$ & 1.4(1) & 1.7(2) & 2.4(4) \\ 
\hline 
$d=3$ & 2.68(9) & 8.2(4) & 22(2) \\ 
\hline 
$d=4$ & 3.79(3) & 24(1) & 91(4) \\ 
\hline 
$d=5$ & 4.7(1) & 30(1) & 141(6) \\ 
\end{tabular} 
\caption{Improvement factors $\tau_{\text{int},i}^{(\mathcal{N})}/\tau_{\text{int},j}^{(\mathcal{N})}$ by changing from the P-S to the B-S type worm algorithm, B-S type to the irreversible worm algorithm, and P-S to the irreversible worm algorithm on the $d$ dimensional torus}
\label{table:torus_improvements}
\end{table}

\subsection{Complete graph}
\label{subsec:complete_graph}
\begin{figure}[htb]
            \centering
\includegraphics[scale=0.35]{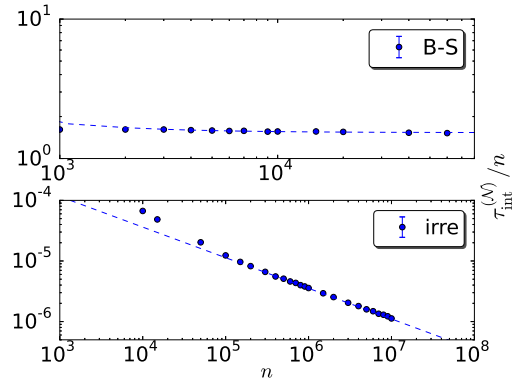}
\caption{Finite-size scaling of $\tau_{\text{int}}^{(\mathcal{N})}/n$ for the B-S type worm algorithm (upper panel) and lifted worm algorithm (lower panel) on the complete graph with $n$ vertices. The dashed lines correspond to the fits in Sec.~\ref{subsec:complete_graph}.}
\label{fig:int_auto_time_bond_irre}
\end{figure}

\begin{figure}[htb]
            \centering
\includegraphics[scale=0.35]{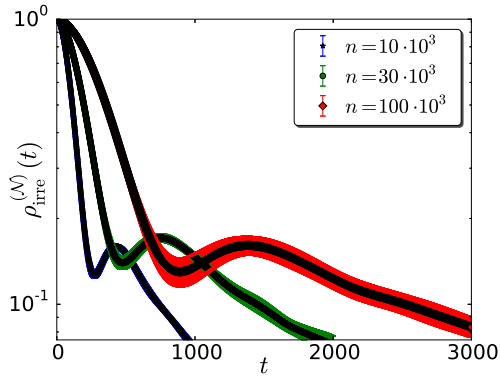}
\caption{Normalized autocorrelation function $\rho_{\rm irre}^{(\mathcal{N})}(t)$ ($t$ in MC hits) for the lifted worm algorithm on the complete graph with $n$ vertices. As stated in Sec.~\ref{subsec:complete_graph}, $\rho_{\rm irre}^{(\mathcal{N})}(t)$ is well described by the ansatz in Eq.~\eqref{autocorrelation_function_compl_graph_irreversible}.}
\label{fig:auto_function_bond_irre_compl_graph}
\end{figure}
We will now study the integrated autocorrelation time of the B-S type and lifted worm algorithm on the complete graph with $n$ vertices. Our main finding is that $\mathcal{N}$ exhibits critical speeding-up for the lifted process. Figure~\ref{fig:int_auto_time_bond_irre} shows $\tau_{\rm int}^{(\mathcal{N})}/n$ for both algorithms where $\tau_{\rm int}^{(\mathcal{N})}$ is measured in MC hits. The fitting ansatz $An^z+B$ for $\tau_{\rm int,irre}^{(\mathcal{N})}$ where $A,B,z$ are free parameters leads to $z=0.51(1)$ by discarding $n<2\cdot 10^5$. Thus, we have $\tau_{\rm int,irre}^{(\mathcal{N})} / n \sim n^{-1/2}$. Note that we had to choose large windowing parameters in agreement to our findings on high-dimensional tori. For the corresponding reversible counterpart (B-S type worm algorithm), it follows immediately from general arguments~\cite[Cor. 9.2.3]{1996_madras} that the integrated autocorrelation time satisfies a Li-Sokal type bound $\tau_{\rm int,\text{B-S}}^{(\mathcal{N})} \ge \text{const} \times \text{Var}(n)$ where $\text{const}>0$. One can, furthermore, calculate that $\lim_{n \to \infty} \frac{\text{Var}(n)}{n} = \frac{9}{4}-\frac{24\Gamma(5/4)^4}{\pi^2}$ leading to $\tau_{\rm int,\text{B-S}}^{(\mathcal{N})}/n \ge \text{const}$.  Thus, lifting improves the dynamic critical exponent on the complete graph. Our numerics suggest that this bound is sharp, i.e.~$\tau_{\rm int,\text{B-S}}^{(\mathcal{N})} / n \sim \text{const}$. More precisely, our fitting ansatz $An^z+B$ for $\tau_{\rm int,\text{B-S}}^{(\mathcal{N})}$ leads to $z= 1.00(2)$ by discarding $n<3000$.\par
Figure~\ref{fig:auto_function_bond_irre_compl_graph} shows the autocorrelation function $\rho_{\rm irre}^{(\mathcal{N})}(t)$ for the lifted process where $t$ is measured in MC hits. Our data suggests that $\rho_{\rm irre}^{(\mathcal{N})}(t)$ is well described by the following equation:
\begin{equation}
\rho_{\rm irre}^{(\mathcal{N})}(t) = (1-\alpha)\exp(-t/\tau_1)+\alpha\cos(\omega t+\phi)\exp(-t/\tau_2).
\label{autocorrelation_function_compl_graph_irreversible}
\end{equation}
Similar ans\"atze were used in other studies on lifting, see~\cite{2011_turitsyn,2016_vucelja}. Our fits lead to $\tau_1 \propto n^{2/3}$, $\tau_2 \propto n^{1/2}$, and $\omega \propto n^{-1/2}$. Moreover, we find that the amplitude of the first term vanishes as $n\to \infty$. We note that the cosine in Eq.~\eqref{autocorrelation_function_compl_graph_irreversible} is motivated by the fact that the eigenvalues of the transition matrix in the lifted process need not be real, unlike the spectrum of reversible chains.\par
Finally, we numerically observe that the average number of consecutive steps  the chain spends in each replica, $\tau_{\rm intra}$, scales as $n^{1/2}$. More precisely, the ansatz $An^z+B$ leads to $z=0.49(1)$ by discarding $n<25$. It is interesting to observe that we have thus identified three time scales, i.e.~$\tau_{\rm int,irre}^{(\mathcal{N})},~\tau_2,~\tau_{\rm intra}$, which scale as $n^{1/2}$.

\section{Discussion} 
\label{sec:discussion} 
We constructed an irreversible MCMC worm algorithm for the zero-field ferromagnetic Ising model via the lifting technique. Since it is not obvious how to lift the standard P-S worm algorithm to generate ballistic motion for the number of occupied edges, we first constructed an alternative worm algorithm with different proposals (B-S type) and lifted this chain. We emphasize that this construction can also be used to design lifted worm algorithms for other important models in statistical mechanics such as the $XY$ model. The $XY$ model shares many universal properties of the Bose-Hubbard model~\cite{2015_svistunov} which is actively studied in ultracold atom physics. We also emphasize that the lifted worm algorithm for the Ising model can be implemented by changing only a few code lines in the B-S type worm algorithm.\par 
We studied the dynamical critical behavior of the number of occupied edges $\mathcal{N}$ on both the complete graph and toroidal grids in various dimensions. On the complete graph we numerically established $z_{\rm int,irre,\mathcal{N}}\approx -0.5$ for the lifted worm algorithm, while we found $z_{\rm int,\text{B-S},\mathcal{N}}\approx0$ for the corresponding reversible counterpart. It is interesting to observe that the mixing time of the Swendsen-Wang algorithm scales as $t_{\rm mix}/n \sim n^{1/4}$~\cite{2009_long}. On the torus we numerically established that the lifted chain leads to a constant improvement with larger improvements for higher dimensions. Note that even though we did not find a dynamic improvement for $d\le 5$, it is an open question if $\mathcal{N}$ exhibits critical speeding-up on higher dimensional tori. We, finally, note that a recent study~\cite{2017_suwa} constructed an alternative irreversible worm algorithm which is based on the so-called geometric allocation approach~\cite{2010_suwa}. Similar to our findings, this irreversible worm algorithm does not improve the dynamical critical exponent for $\mathcal{N}$ on the three-dimensional torus, but leads to a significant constant improvement.\par 

\begin{acknowledgments} 
We would like to thank Zongzheng Zhou for fruitful discussions. E. M. E. is grateful to Nikolaos G. Fytas for fruitful discussions regarding various finite-size scaling ans\"atze. This work was supported under the Australian Research Council’s Discovery Projects funding scheme (Project Number DP140100559). It was undertaken with the assistance of resources from the National Computational Infrastructure (NCI), which is supported by the Australian Government. Furthermore, we would like to acknowledge the Monash eResearch Centre and eSolutions-Research Support Services through the use of the Monash Campus HPC Cluster. Y. Deng thanks the National Natural Science Foundation of China for their
support under Grant No. 11625522 and the Fundamental Research Funds for the Central Universities under Grant
No. 2340000034. Y. Deng also thanks the support from MOST under Grant No. 2016YFA0301600.
\end{acknowledgments}

\appendix
\section{Estimation with the Madras-Sokal automatic windowing algorithm and suppressed slow modes\label{ap:sokal} } 
\begin{figure}[htb]
\includegraphics[width=0.9\columnwidth]{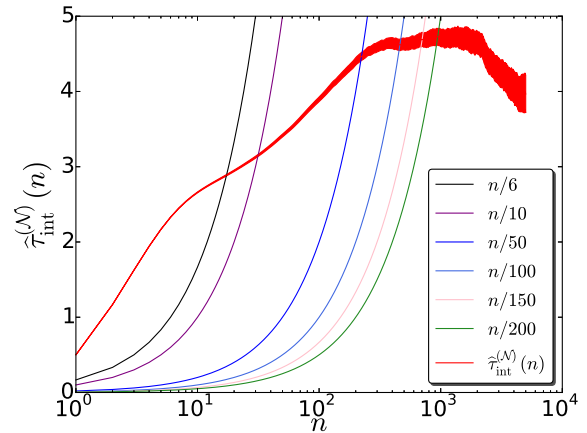}
\caption{Dependence on the cut-off time $n$ of the truncated integration autocorrelation time $\widehat{\tau}_{\rm int}^{(\mathcal{N})}(n)$ defined in \ref{eq:def_tau_int_truncated} for the lifted worm algorithm in five dimensions and $L=56$ at criticality. The solid red line shows $\widehat{\tau}_{\rm int}^{(\mathcal{N})}(n)$ averaged over 100 independent runs, whereas the surrounding region corresponds to $\pm 1 \sigma$. One clearly sees how $n<100$ underestimates $\widehat{\tau}_{\rm int}^{(\mathcal{N})}(n)$ and much larger $n$ are required to capture contributions corresponding to the suppressed mode.}
\label{fig:tauint_cparamchoice}
\end{figure}

A widely used numerical method to estimate the integrated autocorrelation time of a weakly stationary time series~\cite{1981priestley} is to consider a truncated version of~\eqref{eq:def_tau_int}, i.e.~
\begin{equation}
\widehat{\tau}_{{\rm int}}^{(\mathcal{N})}(n) := \frac{1}{2} + \sum_{t=1}^{n} \rho^{(\mathcal{N})}(t).
\label{eq:def_tau_int_truncated}
\end{equation}
The reason for truncating the time series is well known~\cite{1981priestley,1996_sokal}: The standard deviation of ~\eqref{eq:def_tau_int_truncated} is ${\rm O}(\sqrt{n/M})$, and, hence, (for fixed $M$) grows with $n$. The choice of $n$ is, thus, a compromise between bias and standard-error. The Madras-Sokal automatic windowing method determines $n$ self-consistently as the smallest positive integer $n$  that fulfils (numerically)\par
$$
c \tau_{\rm int}^{(\mathcal{N})}(n) \leq n
$$
where $c$ is a free real parameter, typically manually chosen to lie in $[6,10]$, c.f.~\cite{1996_sokal} for a justification of this particular choice of $c$.\\
While analysing the time series for $\mathcal{N}$ for the lifted worm algorithm in $d\geq 4$ we made the observation that the choice $c\in [6,10]$ significantly underestimates the integrated autocorrelation time. Intuitively, what we implement with~\eqref{eq:def_tau_int_truncated} is a `discrete integration' which we utilise to extract the decay constant (the analogue of the relation  $\int_{0}^{\infty} dx~e^{-x/\tau} = \tau$ in a continuous setting). However, in our setting we only obtain a significant contribution from $\alpha_1 e^{-t/\tau_1}$ to \eqref{eq:def_tau_int_truncated} when including lags for which the contribution of $\alpha_2 e^{-t/\tau_2}$ to $\rho^{(\mathcal{N})}(t)$ is too small to be separated from noise. The combination of small amplitude and large time scale, thus, requires adjusting $c$. In Figure~\ref{fig:tauint_cparamchoice} we illustrate this for the particular choice of $L=56$ of the lifted worm algorithm in $d{=}5$.

\section{Least square fitting a weighted exponential ansatz \label{ap:appenfit}}
While fitting the normalised autocorrelation function for $\mathcal{N}$ in dimensions four and five we made the observation that it
is numerically very hard to obtain reliable estimates of the parameters involved in a functional model of the two-time scale behavior using the method of least squares. We, therefore, manually analysed the least square conditions. As result of this we were able to reduce the free parameters which we need to numerically minimise from $4$ to $2$. This significantly improved the numerical precision and stability of the fitting procedure. To this end, our model has free parameters $\alpha_1>0, \alpha_2>0,1>\lambda_1>0, 1>\lambda_2>0$ and we would like to minimise the sum of the first $T_{\max}$ squared residuals 
$$
r_k^2 \equiv \Big[\alpha_1 \lambda_1^k + \alpha_2 \lambda_2^k - \rho_\mathcal{N}(k)\Big]^2,
$$
that is our goal is to minimise the function $\Theta$ defined as
$$
\Theta(\alpha_1,\alpha_2,\lambda_1,\lambda_2) \equiv \sum_{k=0}^{T_{\max}-1}r_k^2
$$

A necessary condition for the minimum is 
$$
\frac{\partial \Theta}{\partial \alpha_1}
= 
\frac{\partial \Theta}{\partial \alpha_2}
= 
\frac{\partial \Theta}{\partial \lambda_1}
= 
\frac{\partial \Theta}{\partial \lambda_2} = 0
$$
We start by evaluating the partial derivatives w.r.t. $\alpha_1$ and $\alpha_2$ and a consequence of this obtain the following set of two linear equations
\begin{eqnarray*}
\alpha_1 P(\lambda_1)  + \alpha_2 S(\lambda_1,\lambda_2) &=& Q(\lambda_1), \\
\alpha_1 S(\lambda_1,\lambda_2) + \alpha_2 P(\lambda_2) &=& Q(\lambda_2).\\
\end{eqnarray*}
Here we defined the following  three polynomials over $(0,1)$ or $(0,1)^2$:
\begin{eqnarray*}
P(\lambda) &\equiv& \sum_{k=0}^{T_{\max} - 1}\lambda^{2k} =\frac{\lambda^{2T_{\max}} - 1}{\lambda^2 - 1} \\
Q(\lambda) &\equiv& \sum_{k=0}^{T_{\max} - 1}\lambda^{k}\rho_\mathcal{N}(k) \\
S(\lambda_1,\lambda_2) &\equiv& \sum_{k=0}^{T_{\max} - 1}\lambda_1^{k}\lambda_2^{k}
\end{eqnarray*}
Note that $S$ is symmetric in $\lambda_1, \lambda_2$. The above set of linear equations has the unique solution:
$$
\alpha_1(\lambda_1, \lambda_2) = \frac{S(\lambda_1,\lambda_2)Q(\lambda_2) - Q(\lambda_1)P(\lambda_2)}{S(\lambda_1,\lambda_2)^2 - P(\lambda_1)P(\lambda_2)}
$$
and $\alpha_2(\lambda_1,\lambda_2)$ is obtained by replacing $1\rightarrow 2, 2\rightarrow 1$ in the r.h.s. above. Thus, this allows us to substitute $\alpha_1,\alpha_2$ as a function of $\lambda_1,\lambda_2$ in $\Theta$ and minimise $\Theta$ only w.r.t. to $\lambda_1,\lambda_2$, with the constraints that $\lambda_1,\lambda_2 \in (0,1)$. We remark, that completely analogous arguments can be used to show that a similar improvement can be achieved for a general ansatz $\sum_{j=1}^{n} \alpha_j \lambda_j^t$, where one reduces the parameters to numerically minimise from $2n$ to $n$. The calculations done here generalise the method very recently proposed in \cite{tauint_paper} for a single exponential ansatz.



\begin{thebibliography}{99}

\bibitem{2016_goodfellow} I. Goodfellow, Y. Bengio and A. Courville, \textit{Deep Learning}, MIT press, (2016).
\bibitem{2016_efron} B. Efron, T. Hastie, \textit{Computer Age Statistical Inference}, Cambridge University Press, Volume 5, (2016).
\bibitem{1953_metropolis} N. Metropolis, A. W. Rosenbluth, M. N. Rosenbluth, A. H. Teller, E. Teller, \textit{Equations of state calculations by fast computing machines}, J. Chem. Phys. 21, 1087 (1953).
\bibitem{2008_janke} W. Janke, \textit{Monte Carlo Methods in Classical
Statistical Physics}, in Computational Many-Particle Physics, Lect. Notes Phys., Vol. 739, Springer-Verlag, Berlin (2008).
\bibitem{2001_prokofev} N.Prokof'ev, B. Svistunov, \textit{Worm Algorithms for Classical Statistical Models}, Phys. Rev. Lett. 87, 160601 (2001).
\bibitem{2006_prokofev} M. Boninsegni, N. Prokof'ev, B. Svistunov, \textit{Worm Algorithm for Continuous-space Path Integral Monte Carlo Simulations}, Phys. Rev. Lett. 96, 070601 (2006).
\bibitem{1989_wolff}U. Wolff, \textit{Collective Monte Carlo Updating for Spin Systems}, Phys. Rev. Lett., 62 (4): 361 (1989)
\bibitem{1987_swendsen} R. H. Swendsen, J.-S. Wang, \textit{Nonuniversal critical dynamics in Monte Carlo simulations}, Phys. Rev. Lett., 58(2): 86–88 (1987)
\bibitem{2007_deng} Y. Deng, T. M. Garoni, A. Sokal, \textit{Dynamic critical behavior of the worm algorithm for the Ising model}, Phys.Rev.Lett. 99, 110601 (2007).
\bibitem{2009_wolff} U. Wolff, \textit{Simulating the all-order strong coupling expansion I: Ising model demo}, Nucl. Phys. B, 810, 491 (2009).
\bibitem{2016_colle} A. Collevecchio, T. M. Garoni, T. Hyndman, D. Tokarev, \textit{The worm process for the Ising model is rapidly mixing}, Journal of Statistical Physics 164, 1082-1102 (2016).
\bibitem{2010_suwa}H. Suwa, S. Todo, \textit{Markov Chain Monte Carlo Method without
Detailed Balance}, Phys. Rev. Lett. 105, 120603 (2010).
\bibitem{2010_hayes} T. P. Hayes, A. Sinclair, \textit{Liftings of Tree-Structured Markov Chains}, Conference: Approximation, Randomization, and Combinatorial Optimization. Algorithms and Techniques, 13th International Workshop, APPROX 2010, and 14th International Workshop, RANDOM 2010, Barcelona, Spain, September 1-3, (2010).
\bibitem{2011_turitsyn}K. Turitsyn, M. Chertkov, M. Vucelja, \textit{Irreversible Monte Carlo algorithms for efficient sampling}, Physica D 240, 410-414 (2011).
\bibitem{2013_sakai} Y. Sakai, K. Hukushima, \textit{Dynamics of one-dimensional Ising model without detailed balance condition}, J. Phys. Soc. Japan 82, 064003-1-8 (2013).
\bibitem{2011_fernandes} H. C. M. Fernandes, M. Weigel, \textit{Non-reversible Monte Carlo simulations of spin models}, Computer Physics Communications 182, 1856 (2011).
\bibitem{2009_bernard}E. Bernard, W. Krauth, D. Wilson, \textit{Event-chain Monte Carlo algorithms for hard-sphere systems}, Phys. Rev. E 80, 056704 (2009).
\bibitem{2013_engel} M. Engel, J. A. Anderson, S. C. Glotzer, M. Isobe, E. P. Bernard, W. Krauth, \textit{Hard-disk equation of state: First-order liquid-hexatic transition in two dimensions with three simulation methods}, Phys. Rev. E 87, 042134 (2013).
\bibitem{2015_kapfer} S. C. Kapfer, W. Krauth, \textit{Soft-disk melting: From liquid-hexatic coexistence to continuous transitions}, Phys. Rev. Lett. 114, 035702 (2015).
\bibitem{2015_michel}M. Michel, J. Mayer, W. Krauth, \textit{Event-chain Monte Carlo for
classical continuous spin models}, EPL 112, 20003 (2015).
\bibitem{2015_nishikawa} Y. Nishikawa, M. Michel, W. Krauth, K. Hukushima, \textit{Event-
chain algorithm for the Heisenberg model: Evidence for $z \sim 1$ dynamic scaling}, Phys. Rev. E 92, 063306 (2015).
\bibitem{2017_hu} H. Hu, X. Chen, Y. Deng, \textit{Irreversible Markov chain Monte Carlo algorithm for self-avoiding walk}, Frontiers of Physics, 12:120503, (2017).
\bibitem{2001_diaconis} P. Diaconis, S. Holmes, R. M. Neal, \textit{Analysis of a nonreversible Markov chain sampler}, Ann. Appl. Probab. Volume 10, Number 3, 726-752 (2000).
\bibitem{2013_diaconis}P. Diaconis, \textit{Some things we've learned (about Markov chain Monte Carlo)}, Bernoulli 19 1294–1305, (2013).
\bibitem{2016_vucelja} M. Vucelja, \textit{Lifting - A nonreversible Markov chain Monte Carlo algorithm}, American Journal of Physics 84, 958 (2016).
\bibitem{2007_deng_2} Y. Deng, T. M. Garoni, A. D. Sokal, \textit{Critical Speeding-Up in the Local Dynamics of the Random-Cluster Model}, Phys. Rev. Lett. 98 230602, (2007).
\bibitem{2011_pomeau} Y. Pomeau, M. Le Berre, \textit{Critical speed-up vs critical slow-down: a new kind of relaxation oscillation with application to stick-slip phenomena}, arXiv:1107.3331, (2011).
\bibitem{2014_lundow} P. H. Lundow, K. Markstr\"om, \textit{Finite size scaling of the 5D Ising model with free boundary conditions}, Nucl. Phys. B, vol 889, p249 (2014).
\bibitem{2014_wittmann}M. Wittmann, A. P. Young, \textit{Finite-size scaling above the upper critical dimension}, Phys. Rev. E 90, 062137 (2014).
\bibitem{2016_flora} E. Flores-Sola, B. Berche, R. Kenna, M. Weigel, \textit{Role of Fourier Modes in Finite-Size Scaling above the Upper Critical Dimension}, Phys. Rev. Lett. 116, 115701 (2016).
\bibitem{2017_grimm} J. Grimm, E. M. El\c{c}i, Z. Zhou, T. M. Garoni, Y. Deng, \textit{Geometric explanation of anomalous finite-size scaling in high dimensions}, Phys. Rev. Lett. 118, 115701 (2017).
	
	
\bibitem{1979_thompson} C. J. Thompson, \textit{Mathematical Statistical Mechanics}, Princeton University Press (1979).
\bibitem{1985_berretti} A. Berretti, A. D. Sokal, \textit{New Monte Carlo method for the self-avoiding walk}, J. Stat. Phys. 40, 483 (1985).

\bibitem{2014_landau} D. P. Landau, K. Binder, \textit{A guide to Monte Carlo simulations in statistical physics}, Cambridge university press, (2014).

\bibitem{1982_baxter} R. J. Baxter, Exactly solved models in statistical mechanics, London: Academic Press Inc. [Harcourt Brace Jovanovich Publishers] (1982).
\bibitem{2003_deng} Y. Deng, H. W. J. Bl\"ote, \textit{Simultaneous analysis of several models in the three-dimensional Ising universality class}, Phys. Rev. E 68, 036125 (2003).
\bibitem{2012_lundow} P. H. Lundow, K. Markstr\"om, \textit{Critical behaviour of the Ising model on the 4-dimensional lattice}, Phys. Rev. E 80, 031104 (2009).
\bibitem{2015_young} P. Young, \textit{Everything You Wanted to Know About Data Analysis and Fitting but Were Afraid to Ask}, SpringerBriefs (2015).
	\bibitem{1996_sokal} A. D. Sokal, \textit{Monte Carlo Methods in Statistical Mechanics: Foundations and New Algorithms}, Lectures at the Carg\`ese Summer School on `Functional Integration: Basics and Applications' (1996).
	\bibitem{2011_liu} Q. Liu, Y. Deng, and T. M. Garoni, \textit{Worm Monte Carlo study of the honeycomb-lattice loop model}, Nucl. Phys. B 846, 283-315 (2011).
	\bibitem{20144disinglundow}  P. H. Lundow and K Markstr\"om, \textit{Critical behavior of the Ising model on the four-dimensional cubic lattice}, Phys. Rev. E 80, 031104 (2009).
	
	
\bibitem{1996_madras} N. Madras, G. Slade, \textit{The Self-Avoiding Walk}, Birkh\"auser, Boston (1996).

\bibitem{2015_svistunov} B. Svistunov, E. Babaev, N. Prokof’ev, \textit{Superfluid
States of Matter}, CRC Press, Boca Raton (2015).
\bibitem{2009_long} Y. Long, \textit{Mixing Time of the Swendsen-Wang Dynamics on the Complete Graph and Trees}, Ph.D. Thesis, UC Berkeley (2009).
\bibitem{2017_suwa} H. Suwa, \textit{Directed Worm Algorithm with Geometric Optimization}, arXiv:1703.03136 [cond-mat.stat-mech] (2017).
\bibitem{tauint_paper} Y. Fang, Y. Cao and R. D. Skeel, \textit{Quasi-Reliable Estimates of Effective Sample Size}, arXiv:1705.03831, (2017).
\bibitem{1981priestley} M. B. Priestley, Spectral analysis and time series: Academic Press Inc. (1981) 
\end{thebibliography}
\end{document}